\documentstyle[11pt,newpasp,twoside,epsf]{article}
\def\edcomment#1{\iffalse\marginpar{\raggedright\sl#1\/}\else\relax\fi}
\reversemarginpar

\begin{document}
\title{The Far Infrared and Submillimeter Diffuse Extragalactic Background} 
\author{M. G. Hauser}
\affil{Space Telescope Science Institute, 3700 San Martin Drive, Baltimore, Maryland 21218, USA}

\begin{abstract}
The cosmic infrared background (CIB) radiation was a long-sought fossil of energetic processes associated with structure formation and chemical evolution since the Big Bang.    The COBE Diffuse Infrared Background Experiment (DIRBE) and Far Infrared Absolute Spectrophotometer (FIRAS) were specifically designed to search for this background from 1.25~$\mu$m to millimeter wavelengths.  These two instruments provided high quality, absolutely calibrated all-sky maps which have enabled the first detections of the CIB, initially at far infrared and submillimeter wavelengths, and more recently in the near infrared as well.  The aim of this paper is to review the status of determinations of the CIB based upon COBE measurements.   The results show that the energy in the CIB from far infrared to millimeter wavelengths is comparable to that in the integrated light of galaxies from UV to near infrared wavelengths: the universe had a luminous but dusty past.  On the assumption that nucleosynthesis in stars is the energy source for most of this light, the results also imply that 1--8\% of cosmic baryons has been converted to helium and heavier elements in stars.  The integrated background light from UV to millimeter wavelengths, 60--120 \mbox{nW m$^{-2}$ sr$^{-1}$}, is about 10\% of that in the cosmic microwave background.  Current knowledge of the CIB provides significant new constraints on models of the history of star formation and galaxy evolution.
\end{abstract}

\section{Introduction}

Cosmic infrared background (CIB) radiation is expected to arise from 
the cumulative emissions of pregalactic, protogalactic, and evolved 
galactic systems.  It has long been recognized that measurement of the CIB 
will provide new insight into energetic processes associated with structure  
formation and chemical evolution following the
decoupling of matter from the cosmic microwave background (CMB)
radiation (Partridge \& Peebles 1967; Harwit 1970; Bond, Carr, \&
Hogan 1986, 1991; Franceschini et al.\ 1991, 1994; Fall, Charlot, \&
Pei 1996).  In this paper, the CIB is defined to be 
extragalactic radiation, exclusive of the CMB, in the wavelength range 1--1000~$\mu$m. 

Sources of cosmic radiant energy include nuclear processes such as
nucleosynthesis in stars; gravitational processes, such as accretion of matter
onto black holes; and decaying unstable particles remaining from the Big Bang.
Though the primary radiant energy from such processes may not emerge at infrared 
wavelengths, the combined effects of the cosmic redshift and absorption of some fraction of the primary 
radiations by dust with re-emission by the dust at long wavelengths will shift much of the energy into 
the infrared.  Hence, the infrared background is expected to be uniquely informative about cosmic history.  

Sky brightness measurements with instruments on NASA's Cosmic Background Explorer (COBE) 
mission have provided the first definitive detections of the CIB.  In this paper I focus 
on the {COBE} measurements, which initially provided detections only at far infrared and submillimeter wavelengths.  I will also describe recent results of analyses based upon COBE data that have provided likely detections in the near infrared and restrictive limits in the mid-infrared, since these results are not covered substantially elsewhere at this conference.  I then briefly describe some of the implications of these results.  A comprehensive review paper on the infrared background and its implications is in preparation (Hauser \& Dwek 2001).  Since energy distributions are of primary interest, sky brightness measurements are reported as 
$\nu$I$_\nu$ in units of \mbox{nW m$^{-2}$ sr$^{-1}$}, where I$_\nu$ is the spectral intensity at frequency $\nu$.  Conversion to I$_\nu$ can be accomplished using the relation $\nu$I$_\nu$(\mbox{nW m$^{-2}$ sr$^{-1}$}) = [3000/$\lambda\,(\mu$m)]\,I$_\nu$(MJy/sr).

\section{The COBE Mission}

Measurement of the CIB has proven to be extraordinarily difficult.  The only observational signature of the CIB is an isotropic background arising external to the Galaxy.  There is no distinctive, predictable spectral signature of the CIB, since many sources of luminosity can contribute, and their distribution in space and time, as well as the character and distribution of cosmic dust, will determine the observed spectral energy distribution. 
Direct observation of the CIB is impeded by the many foreground contributors to the infrared sky brightness at all
wavelengths, several of them quite bright.  A further challenge is the need for absolute photometry, that is, measurements relative to a well-established zero brightness level.  Instrument self-emission, stray light, and electronic offset signals must be eliminated or accurately determined.

The COBE mission was designed to accomplish the best possible search for the CIB from our location in the cosmos.  It was recognized that such measurements required cryogenic instruments operated above the atmosphere for a substantial period of time so that potential sources of systematic errors in the absolute brightness measurements could be identified and evaluated.  The approach was to obtain maps of high photometric quality over the whole sky and over a broad wavelength range so that foreground sources could be identified by their spatial and spectral signatures, and that uncertainties associated with discriminating them from the CIB could be minimized.  
The COBE spacecraft carried two instruments contributing to these results (Boggess et al.\ 1992).  The Diffuse Infrared Background Experiment (DIRBE) was designed primarily to search for the CIB from 1.25 to 240~$\mu$m.  The Far Infrared Absolute Spectrophotometer (FIRAS) was designed primarily to make a definitive measurement of the spectrum of the CMB, and to extend the search for the CIB to millimeter wavelengths.  All aspects of the COBE mission, from the instruments to the spacecraft, orbit, sky scan strategy, and data processing, were designed to optimize the ability to make these difficult diffuse background measurements.

The DIRBE instrument was an absolute photometer with 10 broad photometric bands
at 1.25, 2.2, 3.5, 4.9, 12, 25, 60, 100, 140, and 240~$\mu$m.  It was designed to enable detection of CIB levels as faint as $\nu I_\nu \sim 1$ \mbox{nW m$^{-2}$ sr$^{-1}$}.  The DIRBE instrument is described by Silverberg et al.\ (1993), and a summary of the DIRBE investigation and its initial results is provided by Hauser et al.\ (1998).  Additional detailed information is given in the COBE DIRBE Explanatory Supplement (1997).   The DIRBE was designed for extremely strong stray light rejection, employing an off-axis Gregorian telescope, with a pupil stop, multiple field stops, and extensive internal and external baffling.  The stray light was demonstrated to be less than 1 \mbox{nW m$^{-2}$ sr$^{-1}$} at all wavelengths.  The instrument contained an internal cold chopper operating at 32 Hz, which allowed continual measurement of the sky brightness relative to that of a stable, cold internal beam stop.  It had a full beam cold shutter, closed typically five times per orbit to establish the zero point of the photometric scale by allowing measurement of the instrumental radiative and electronic zero point offsets.  These offsets were stable, and the uncertainty in the determination of the offsets was 1 \mbox{nW m$^{-2}$ sr$^{-1}$} or less from 1.25 to 100~$\mu$m, and 
was 5 (2) \mbox{nW m$^{-2}$ sr$^{-1}$}~at 140 (240)~$\mu$m respectively.  

The DIRBE instantaneous field of view was 0.7$^\circ$ x 0.7$^\circ$, a compromise between ability to discriminate stars and ability to map the whole sky with high redundancy every six months.  Redundancy was important because it allowed monitoring of the annual variations in apparent sky brightness in all directions due to the Earth's motion within the interplanetary dust (IPD) cloud.  This variation is a unique signature of the IPD contribution to the signal.  The sensitivity (1$\sigma$) of the instrument from 1.25 to 100~$\mu$m was better 
than 1 \mbox{nW m$^{-2}$ sr$^{-1}$} per field of view averaged over the ten months of cryogenic operation, and was 33 (11) \mbox{nW m$^{-2}$ sr$^{-1}$} at 140 (240)~$\mu$m respectively.  The instrument gain stability was excellent, and was monitored on short time scales using internal stimulators.  Repeated observations of stable celestial sources provided photometric closure over the sky, and assured reproducible photometry to $\sim$ 1\% or better for the duration of the mission.  Calibration of the DIRBE flux scale was accomplished from scans of a few isolated infrared sources of known brightness.

The FIRAS instrument was a Fourier transform spectrometer in the form of a polarizing Michelson interferometer, providing extremely precise spectral comparison of the sky brightness with that of a very accurate full beam blackbody calibrator at wavelengths from 100~$\mu$m to 1 cm.  The instrument and calibration are discussed by Mather et al.\ (1993), 
Fixsen et al.\ (1994), Mather et al.\ (1999), and the COBE FIRAS Explanatory Supplement (1997).  The FIRAS light collector was a Winston cone with a flared horn, providing a 7$^\circ$ diameter field of view and low sidelobe response over the broad FIRAS spectral range.  The sensitivity (1$\sigma$) of the instrument from 500~$\mu$m to 3 mm was 0.8 \mbox{nW m$^{-2}$ sr$^{-1}$} per field of view averaged over the ten months of cryogenic operation.  The photometric calibration errors (1$\sigma$) associated with the precision blackbody calibrator were typically 0.02 MJy/sr, corresponding to $\nu I_\nu \sim 0.3\ (0.1) $ \mbox{nW m$^{-2}$ sr$^{-1}$} at 200 (600)~$\mu$m wavelength respectively.  Over the course of the COBE mission, the FIRAS obtained superbly calibrated absolute spectral maps of almost the whole sky.  
Using FIRAS data, Fixsen et al.\ (1996) and Mather et al.\ (1999) showed that the rms deviation of the CMB spectrum from that of a $(2.725\ \pm\ 0.002)$ K blackbody at wavelengths longer than 476~$\mu$m was less than 50 parts per million of the peak CMB brightness.  Accurate knowledge of the CMB is necessary to search for the CIB at submillimeter wavelengths.

Fixsen et al.\ (1997) demonstrated that the zero point and gain calibrations of the DIRBE and FIRAS photometric scales are consistent within the quoted uncertainties of each and the systematic uncertainties of the comparison.  Since the FIRAS systematic calibration uncertainties are smaller than those of the DIRBE, these results can be used to make small systematic corrections to the DIRBE 140 and 240~$\mu$m measurements.  Hauser et al.\ (1998) discussed the effect of using the FIRAS calibration on the DIRBE 140 and 240~$\mu$m CIB detections, but DIRBE-based results in the literature, including this paper, otherwise use the DIRBE calibration.

Hence, the DIRBE and FIRAS instruments provided extensive, consistent, high quality photometric data on which to base a search for the CIB.

\section{The Submillimeter Background}

The COBE data have provided detections of the CIB at some wavelengths and limits elsewhere within the spectral range of the two instruments.  Since foreground sources and the CMB dominate the observed sky brightness in all directions throughout the \mbox{1--1000~$\mu$m} spectral range, all of these results require discrimination and removal of non-CIB contributions.  Assessment of the reality of a potential CIB measurement therefore requires careful analysis of the random and systematic uncertainties in the residuals from the measured sky brightness.  The residual must also be demonstrated to be isotropic, and not likely to arise from any unmodeled foreground.  

Quantitative results from all of the investigations described here are listed in Table~1.  Error bars for reported detections are 1$\sigma$.  Upper and lower limits are shown at the 95\% CL, with actual values and 1$\sigma$ uncertainties shown in parentheses.  The third column indicates whether some degree of isotropy was demonstrated.  The fourth column indicates whether the investigators claimed a detection (yes), a possible detection (tentative), or a limit (no).  The designation ``no'' is also used for any case where the result is less than 3 times the quoted uncertainty.  For illustrative purposes, the values shown in bold type in Table~1 are plotted in Figure~1.  


\begin{table}

\caption{COBE Diffuse Infrared Background Measurements}

\begin{tabular}{@{}lllll@{}}
\tableline
 $\lambda$ &  $\nu I_{\nu}$ &Isotropy &Detection &  Reference \\
 ($\mu$m) &
 (nW m$^{-2}$ sr$^{-1}$) &
 test passed &
 &
 \\

\tableline

  1.25  & {\bf $<$ 75} (33$\pm$21)         &  no      & no  & Hauser et al.
 1998 \\
  1.25   & $<$ 68 (27$\pm$21)    	 &  no	& no	& Dwek \& Arendt 1998 \\
\vspace{0.1in}
  1.25  & $<$ 57 (27.8$\pm$14.5)      &  no      & no  & Wright 2000  \\

  2.2   & {\bf $<$ 39} (15$\pm$12)         &  no      & no  & Hauser et al.
 1998 \\
  2.2   &   {\bf 23.1$\pm$5.9}        	 &  yes  	& yes	& Wright \& Reese 2000 \\
  2.2   &   22.4$\pm$6.0        	 &  no      & tentative	& Gorjian et al.\ 2000  \\
\vspace{0.1in}
  2.2   &   19.9$\pm$5.3  	 	 &  yes     & yes	&  Wright 2000 \\

  3.5   & {\bf $<$ 23} (11$\pm$6)          &  no      & no  & Hauser et al.
(1998) \\
  3.5   & {\bf 14.4$\pm$4.6}        	 &  no	& tentative	& Dwek \& Arendt 1998 \\
  3.5   &   11.0$\pm$3.3        	 &  no	& tentative	&  Gorjian et al.\ 2000 \\
\vspace{0.1in}
  3.5   & {\bf 12.4$\pm$3.2}        	 &  yes	& yes	&  Wright \& Reese 2000\\

  4.9   & {\bf $<$ 41} (25$\pm$8)          &  no      & no  & Hauser et al.
 1998 \\
\vspace{0.1in}
  4.9   & $<$ 36 (23$\pm$6)    	 &  no	& no	& Dwek \& Arendt 1998 \\

\vspace{0.1in}
  12    & {\bf $<$ 468} (192$\pm$138)      & no       & no  & Hauser et al.
 1998 \\

\vspace{0.1in}
  12--100 & {\bf $<$ 15}			 & no		& no  & Kashlinsky et al.\ 1996b \\

\vspace{0.1in}
  25    & {\bf $<$ 504} (192$\pm$156 )     & no       & no  & Hauser et al.
 1998 \\

  60    & {\bf $<$ 75} (21$\pm$27)         & no       & no  & Hauser et al.
 1998 \\
\vspace{0.1in}
  60    & {\bf 28.1$\pm$7.2}        	 &  yes	& tentative  & Finkbeiner et al.\ 2000
\\

  100   & {\bf $<$ 34} (22$\pm$6)          & no       & no  & Hauser et al.
 1998 \\
  100   & {\bf $>$ 5}  (11$\pm$3)          & no       & no  & Dwek et al.\ 1998
\\
 100    & 23.4$\pm$6.3  	 & no		& yes	& Lagache et al.\ 2000 \\
\vspace{0.1in}
 100    & {\bf 24.6$\pm$8.4}        	 & yes	& tentative	& Finkbeiner et al.\ 2000 \\

  140   & {\bf 32$\pm$6.5}          	 & yes	& tentative	& Schlegel et al.\ 1998 \\
  140   & {\bf 25$\pm$7}             & yes      & yes & Hauser et al.
 1998\\
  140   & $<$ 28 (15.3$\pm$6.4)     & no		& no  & Lagache et
al. 1999 \\
\vspace{0.1in}
  140   & $<$ 47 (24.2$\pm$11.6)     & no		& no  & Lagache et
al. 2000 \\

  240   & {\bf 17$\pm$2}            	 & yes	& tentative  & Schlegel et al.\ 1998 \\
  240   & {\bf 13.6$\pm$2.5}         & yes      & yes & Hauser et al.
 1998\\
  240   & 11.4$\pm$1.9  	 	 & no		& yes   &  Lagache et al.
 1999 \\
\vspace{0.1in}
  240   & $<$ 25 (11.0$\pm$6.9)  	 & no		& no   &  Lagache et al.
 2000 \\

 200-1000 &  {\bf (see Figure~1)} 
   & yes &  tentative & Puget et al.\ 1996\\
 200-1000 & $a \left({\nu \over \nu_0}\right)^k \ \nu B_{\nu}(T)$
      & yes & yes & Fixsen et al.\ 1998$^{(a)}$ \\
\vspace{0.2in}
 200-1000  & $a \left({\nu \over \nu_0}\right)^k \ \nu B_{\nu}(T)$ 
   & yes & yes  & Lagache et al.\ 1999$^{(b)}$ \\
\tableline
\tableline
\vspace{0.05in}
\end{tabular}

$^{(a)}\  a=(1.3\ \pm\ 0.4) \times  10^{-5},\ k=0.64\ \pm\ 0.12,\ T=(18.5\ \pm\ 1.2)\ $K,\ 
$\lambda_0\ =\ 100\ \mu$m 
\newline

$^{(b)}\  a=8.8  \times  10^{-5},\ k=1.4,\ T=13.6\  $K,\ $\lambda_0\ =\ 100\ \mu$m 
\newline
\normalsize

\end{table}

Since the DIRBE team provided extensive foreground models, some of which have been used by other investigators, it is convenient to describe the results and methods of the DIRBE team first.   
The DIRBE team searched for the CIB in the data from all 10 DIRBE wavelength bands, and results based upon the final photometric reduction of the DIRBE data were reported by Hauser et al.\ (1998).  Since there is no direction in the sky devoid of foreground radiation from the solar system and Galaxy, the faintest measured sky brightness at each wavelength is a conservative upper limit on the CIB.  These dark sky limits (95\% CL) are $\nu I_\nu < $ 398, 151, 64, 193, 2778, 2820, 315, 94, 73, and 28 \mbox{nW m$^{-2}$ sr$^{-1}$} at 1.25, 2.2., 3.5, 4.9, 12, 25, 60, 100, 140 and 240~$\mu$m respectively.  While these limits are free of model assumptions, they are too dominated by foreground sources to be of cosmological interest.

After modeling and removing the contributions from all foreground sources, the DIRBE team found isotropic residuals exceeding 3$\sigma$ only at 140 and 240~$\mu$m, as indicated in Table~1.  The DIRBE team further argued that these residuals could not plausibly arise from unmodeled components of the solar system or Galaxy (Dwek et al.\ 1998), and concluded that the CIB had been measured at those wavelengths.  At all other DIRBE wavelengths they reported CIB upper limits (95\% CL) based upon the residuals and estimated uncertainties.  They also determined a lower limit at 100~$\mu$m based on the argument that a thermal source producing the spectrum detected at 140 and 240~$\mu$m could not be fainter than this level at 100~$\mu$m (Dwek et al.\ 1998).

\begin{figure}
\plotone{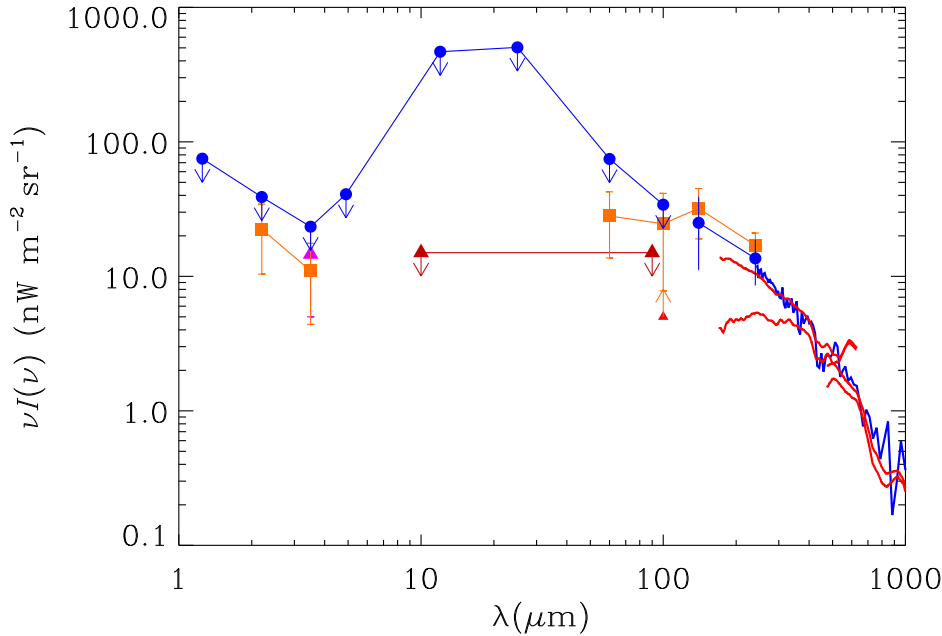}
\caption{Summary of COBE-based CIB measurements.  Detections are shown with 2$\sigma$ uncertainties.  Upper limits are given at the 2$\sigma$ level.  References are listed in Table 1.  The curves at submillimeter wavelengths are from 
Fixsen et al.~1998 (dark wiggly curve from 200--1000~$\mu$m, representing the average of their three approaches); and Puget et al.~1996 (light curves from 150--1000~$\mu$m with break at $\sim$500~$\mu$m: upper curve has no correction for emission correlated with ionized gas, lower curve has the correction).}
\end{figure}

The uncertainties reported by the DIRBE team were dominated by systematic uncertainties in the foreground determination.  This met the mission objective of searching for the CIB to the limits imposed by our astrophysical environment.  Major sources of uncertainty were the stellar foreground model (1.2--3.5~$\mu$m), the interplanetary dust model 
(1.25--100~$\mu$m), and the interstellar medium model (100--240~$\mu$m).  Because DIRBE data can yield improved measurements of the CIB as the foregrounds are better determined, it is worth summarizing how the DIRBE team analysis was done.  

The interplanetary dust contribution presents the most difficult problem except at the longest wavelengths, since it dominates the measured sky brightness from 1--140~$\mu$m even at high galactic and ecliptic latitude (see Figure 2, Hauser et al.\ 1998).  The contribution from the IPD was determined by fitting a parameterized model of the spatial distribution and scattering and radiative properties of the dust cloud to the apparent annual variation of the sky brightness over the whole sky (Kelsall et al.\ 1998).  Though this model was quite successful, reproducing the seasonal variation well and leaving evident map artifacts only at the few percent level, the uncertainties in the small residual differences between the measured brightness and the model are still quite substantial.  Furthermore, it is not possible to determine a unique IPD model from sky brightness measurements from within the cloud.  Kelsall et al.\ considered a number of different shapes for the cloud density distribution which gave comparably good fits to the time dependence, and estimated the model uncertainty by the spread in residual values at high Galactic latitude between these models.  

Arendt et al.\ (1998) described the DIRBE team methods for determining the contribution of Galactic sources.  Discrete bright sources within the Galaxy and the Magellanic Clouds were blanked out of the DIRBE maps.  The contribution from discrete Galactic sources fainter than the DIRBE direct detection limit (about 5th magnitude at 2.2~$\mu$m) was calculated from 1.25 to 25~$\mu$m using a faint source model based on the SKY model of Wainscoat et al.\ (1992).  The SKY  model fits star count data over many wavelengths very well.  

The contribution from the interstellar medium (ISM) was identified by its angular variation over the sky.  The 100~$\mu$m map, after removal of the IPD contribution and an estimated value of the 100~$\mu$m CIB, was used as a template for ISM emission.  The estimate of the 100~$\mu$m CIB was obtained by correlating 100~$\mu$m emission with H~I column density at well-studied dark fields near the North Ecliptic Pole and the Lockman Hole and extrapolating the correlation to zero H~I column density.  This correlation is linear at low column densities.  If all of the Galactic infrared emission were correlated with H~I gas, extrapolation of this correlation to zero column density would yield the extragalactic light.  If there is additional infrared emission associated with other gas components not correlated with H~I, this procedure would not yield accurate results.  Arendt et al.\ (1998) chose these two fields because the molecular gas has also been mapped, and low limits on the contribution from ionized gas could be set using pulsar dispersion measures and H$\alpha$ measurements.  The residual maps at all wavelengths other than 100~$\mu$m, after removal of the IPD contribution and blanking of bright sources, were correlated with the ISM template to determine a global color for each.  The template was then scaled by that color and subtracted from each map.  These final residual maps were tested for positive residuals and isotropy at high galactic latitude where foregrounds are least bright.  Note that an error in the estimated 100~$\mu$m CIB would not have affected the isotropy tests of the residual maps, only the brightness of the residual. The uncertainty in the estimated 
100~$\mu$m CIB was included in estimating the CIB uncertainties at all other wavelengths.  The residual maps at 140 and 
240~$\mu$m showed significant, isotropic residuals in dark regions of the sky, but the most precise determination of the brightness of the residual at these wavelengths was obtained by correlating the emission at these wavelengths with H~I column density in the dark fields in the Lockman Hole and North Ecliptic Pole regions, as described above.  To remove the ISM from the 100~$\mu$m map for purposes of looking for an isotropic residual at 100~$\mu$m, the H~I map of 
Stark et al.\ (1992) was used as the ISM template.  The final 100~$\mu$m residual map did not pass the isotropy tests.

Prior to publication of the DIRBE team results, a tentative identification of the CIB using DIRBE data was reported by Schlegel, Finkbeiner, \& Davis (1998).  In the course of preparing Galactic reddening maps based on the far infrared data from {\it IRAS} and DIRBE, they removed the zodiacal emission contribution using the DIRBE 25~$\mu$m map as a template.  Correlating the residual maps with H~I maps they found a significant constant additional term at 140 and 240~$\mu$m which they identified as a possible measurement of the CIB.  However, they were not certain that the residual constant could not be an instrumental effect. They were therefore not confident that the CIB had been detected, and noted that the results of the DIRBE team would be more definitive.  As Table~1 shows, their tentative values for the CIB are consistent with those of Hauser et al.\ (1998) within the stated uncertainties.

Puget et al.\ (1996) reported the first tentative identification of the CIB.   Analyzing the initial release of FIRAS data, they concluded that there was a residual uniform background from 200~$\mu$m to 2 mm in excess of contributions from the CMB, interplanetary dust, and interstellar dust.  They constructed a simple model for the IPD brightness based upon DIRBE 25 and 100~$\mu$m data and the assumption that the zodiacal emission is only a function of ecliptic latitude.  They extrapolated this model to longer wavelengths assuming $I_\nu \propto \nu^3$.  Since the IPD contribution to the sky brightness at these wavelengths is small, an accurate IPD model is not required. They assumed that the interstellar emission at high galactic latitude and low H~I column density is traced by H~I, and that the submillimeter emission correlated with H~I can be represented by a single temperature medium with a $\nu^2$ emissivity law.  They also made a correction for infrared emission associated with ionized gas not correlated with the H~I, which significantly reduced their residual at the shortest wavelengths \mbox{(200--400~$\mu$m)}.  Their residual maps showed no significant gradients with galactic latitude or longitude, so the residual background was at least approximately isotropic.  Figure~1 shows the residual spectra of Puget et al.\ with and without the ionized gas correction, with a break around 500~$\mu$m due to the shift between the FIRAS high and low frequency channels.  The lower intensity curve includes the correction for Galactic emission from ionized gas. 

A more extensive analysis of the FIRAS data, based upon the final photometric reduction of those data, was presented by Fixsen et al.\ (1998).  They subtracted the CMB contribution and a contribution from interplanetary dust 
using an extrapolation of the DIRBE team model (Kelsall et al.\ 1998) out to \mbox{500~$\mu$m.}   They used three different methods to look for an isotropic residual distinct from emission from the ISM.  One method assumed that the ISM spectrum is the same in all directions, but the intensity is spatially varying.  The second method assumed that the neutral and ionized phases of the ISM are traced by a combination of maps of H~I 21-cm line emission and of C~II (158~$\mu$m) line emission.  The latter was mapped over the sky by FIRAS (Bennett et al.\ 1994).  Each component, including a term proportional to the square of the H~I intensity, was allowed to have a distinct spectrum.  In the third method, they assumed that the ISM emission is traced by a linear combination of the DIRBE ISM maps at 140 and 240~$\mu$m (Arendt et al.\ 1998), each term again having its own spectrum.  Though each of these methods is subject to quite different systematic errors, the three methods yielded a consistent residual isotropic background, providing confidence that this is a robust determination of the submillimeter spectrum of the CIB.  Figure~1 shows the mean spectrum found by Fixsen et al., which is consistent with the DIRBE results of Hauser et al.\ (1998).  Table~1 gives an analytic representation of the mean CIB found by Fixsen et al.\ 

Lagache et al.\ (1999) extended the study of the far infrared emission of the Galaxy at high galactic latitude using FIRAS  data, finding a component of the emission not correlated with H~I emission but which follows a $\csc(b)$ law.  They attributed this emission to the warm ionized medium in the Galaxy.  Subtracting the emission associated with these two gas-phase components from the mean FIRAS spectrum in low H~I column density regions, they obtained the spectrum of the CIB longward of 200~$\mu$m.  The result is consistent, within the uncertainties, with that of Fixsen et al.\ (1998).  An analytic representation of this result is given in Table~1.  They also obtained reduced values of the DIRBE residual in the Lockman Hole region at 140 and 240~$\mu$m.  As shown in Table~1, the 240~$\mu$m value is within 1$\sigma$ of the Hauser et al.\ (1998) value, and the 140~$\mu$m value, which differs by less than 2$\sigma$, is positive by $<3\sigma$ and therefore most confidently provides an upper limit.  

Recently, Lagache et al.\ (2000) extended the study of infrared emission from the warm ionized medium using data from the 
Wisconsin H$\alpha$ Mapper (WHAM) sky survey of Reynolds et al.\ (1998).   They analyzed diffuse emission regions covering about 2\% of the sky at high galactic latitude.  Assuming a constant electron density so they could relate emission measure to H$^{+}$ column density, they found a marginally better correlation of infrared emission at FIRAS resolution with a linear combination of H~I and H$^{+}$ column density than with H~I alone.  Their results suggest that 20--30\% of the far-infrared emission at high galactic latitudes is uncorrelated with H~I gas.  Their resulting values for the CIB at submillimeter wavelengths are consistent with those of Fixsen et al.\ (1998) and Lagache et al.\ (1999).  At 140 and 240~$\mu$m their results are consistent with those of Hauser et al.\ (1998), but with residuals less than 3$\sigma$ positive due to the small sky area analyzed and the uncertainties in the WHAM data.  They obtained a significantly positive residual at 100~$\mu$m similar to that of Hauser et al.\ (1998), but did not demonstrate that this residual was isotropic.

\section{The Far Infrared Background}

Recognizing that the uncertainties in isolating the bright contribution from the IPD are a major obstacle to CIB detection in the far infrared, Finkbeiner, Davis, \& Schlegel (2000) analyzed the DIRBE data using two approaches which avoid the need for a detailed IPD model.  Both methods were based on analysis of the annual variation of dimensionless ratios of the data after removal of the ISM contribution.  The approach of Schlegel et al.\ (1998) was used to determine the ISM contribution.  Subtraction of the ISM contribution from each DIRBE weekly map yielded maps nominally containing only the IPD signal, which varies over the course of the year, and any time-independent signals due to other Galactic contributions and the CIB.  They constructed dimensionless ratios of the brightnesses in opposing directions on the sky, and compared the annual variation or ecliptic latitude variation of these ratios to models containing contributions from interplanetary dust plus a temporally constant background.  Though these methods do not require detailed models of the IPD cloud, they do require assumptions about the spatial symmetries and temporal invariance of the IPD cloud similar to those of the Kelsall et al.\ (1998) model.  In one approach, they analyzed the annual variation of 
[N-S]/[N+S], where N (S) is the brightness toward the North (South) ecliptic pole.  This analysis yielded statistically significant temporally constant backgrounds at 60 and 100~$\mu$m, but does not demonstrate isotropy of that signal.  To explore the ecliptic latitude dependence of the background, they analyzed a second dimensionless ratio based upon brightnesses at 90$^\circ$ solar elongation and symmetric ecliptic latitudes forward and backwards relative to the Earth's direction of motion.  Interpretation of the results of this method does depend modestly on assumptions about the ecliptic latitude dependence of the IPD contribution, i.e., on a model for the IPD cloud at high ecliptic latitudes.  These two methods gave comparable results for the constant backgrounds at 60 and 100~$\mu$m.  Finkbeiner et al.\ tentatively identified these  backgrounds with the CIB (Table~1), but noted that the values are rather high and not consistent with limits implied by the opacity of the intergalactic medium to TeV $\gamma$-rays (Aharonian et al.\ 1999; Samuelson et al.\ 1998).  They concluded that there is not yet a satisfactory explanation for these constant backgrounds.

A distinctly different approach to studying the CIB is to look for fluctuations attributable to the non-uniform spatial distribution of discrete sources contributing to the background.  If one can estimate the expected amplitude of the fluctuations relative to the brightness of the background itself, one can use observations of the fluctuations to determine or set limits upon the CIB.  Of course, measuring fluctuations in the extragalactic background brightness presents similar challenges of discriminating contributions from foreground sources and instrumental effects as those that exist for direct detection of the background.  However, at mid-infrared wavelengths, the dominant IPD foreground is less of an obstacle to fluctuation studies than to total CIB brightness measurements since the IPD emission varies relatively smoothly over the sky.  

Kashlinsky et al.\ (1996a) provided a framework for such analyses.  In order to relate fluctuations to the intensity of the CIB, they evaluated the expected fluctuations from galaxies, finding that the fluctuations on the 0.7$^\circ$ angular scale of the DIRBE beam are expected to be of the order of 5--10\% of the CIB.  Kashlinsky, Mather, \& Odenwald (1996b) analyzed the fluctuations in the DIRBE maps from 12--100~$\mu$m after removing the IPD light using the model of Kelsall et al.\ (1998) and blanking bright discrete sources.  They calculated the zero lag autocorrelation function, $C(0)$, finding $C(0)^{1/2}\ \leq\ (1-1.5)$ \mbox{nW m$^{-2}$ sr$^{-1}$}.  Kashlinsky \& Odenwald (2000) further analyzed the fluctuations in this spectral range, finding a slightly lower limit $C(0)^{1/2}\ \leq\ (0.5-1)$ \mbox{nW m$^{-2}$ sr$^{-1}$}.  These analyses imply upper limits to the 12--100~$\mu$m CIB contributed by sources clustered like galaxies of the order of 10--15 \mbox{nW m$^{-2}$ sr$^{-1}$}, the most restrictive limits based upon infrared measurements in this spectral range (Table~1).  

\section{The Near Infrared Background}
Following the work of Hauser et al.\ (1998), there has been significant progress in direct determination of the CIB in the near infrared window using DIRBE data.  Since the uncertainty in the statistical model of the faint stellar contribution used by Arendt et al.\ (1998) was a major source of uncertainty at these wavelengths, several approaches have been used to reduce that uncertainty.  

Dwek \& Arendt (1998) assumed that the CIB at 2.2~$\mu$m is close to the integrated light from galaxies at this wavelength.  Subtracting the integrated galaxy light and the zodiacal light (Kelsall et al.\ 1998) from the DIRBE \mbox{2.2~$\mu$m} maps yielded a map of starlight at 2.2~$\mu$m (the ISM contribution at this wavelength is negligible).  Using this 2.2~$\mu$m starlight map as a spatial template for starlight at 3.5~$\mu$m, they obtained a significantly positive 3.5~$\mu$m residual, 
$\nu I_\nu = 9.9+0.312[\nu I_\nu(2.2)-7.4] \pm 2.9$ \mbox{nW m$^{-2}$ sr$^{-1}$}, where $\nu I_\nu(2.2)$ is the actual CIB at 2.2~$\mu$m.  They tentatively identified this 3.5~$\mu$m residual as the CIB, though they did not demonstrate that it was isotropic.  They also obtained somewhat improved upper limits on the CIB at 1.25 and 4.9~$\mu$m using the same stellar template (Table~1).

Gorjian, Wright, \& Chary (2000) reduced the stellar foreground uncertainty more directly by measuring all of the stars brighter than 9th magnitude at 2.2 and 3.5~$\mu$m in a dark 2$^\circ$ x 2$^\circ$ field near the north Galactic pole.  They calculated the contribution of fainter stars using the statistical model of Wainscoat et al.\ (1992). The uncertainty in the calculated light from objects below this faint limit is negligibly small, even at high galactic latitude.  They also used a model for the zodiacal light contribution which differed from that of Kelsall et al.\ (1998).  They argued that the Kelsall et al.\ model left a high galactic latitude residual at 25~$\mu$m which is dominated by IPD emission.  The IPD model used by Gorjian et al.\ was similar to that of Kelsall et al.\ in that it required that the apparent annual time variation be reproduced, but it further required that the residual brightness at 25~$\mu$m after zodiacal light removal be constant at a value of zero at high galactic latitude (the ``very strong no zodi principle" of Wright 1997).  After removing the IPD and stellar contributions, Gorjian et al.\ found significant positive residuals at 2.2 and 3.5~$\mu$m which they identified as probable detections of the CIB (Table~1).  Had they used the IPD model of Kelsall et al, their results would have been $\sim$40\% higher than those shown in Table~1, a clear illustration of the uniqueness problem in modeling the zodiacal light.  With a field covering only $\sim$8 DIRBE beams, they did not demonstrate the isotropy of these signals.

Wright \& Reese (2000) compared the histograms of the pixel intensity distributions in the DIRBE 2.2 and 3.5~$\mu$m maps in five fields at high galactic and high ecliptic latitudes with the histograms predicted from the star count model of Wainscoat et al.\ (1992).  The IPD contribution had first been removed from the observations using the model of Gorjian et al.\ (2000). They found that the predicted histograms had the same shape as those observed, but had to be displaced by a constant intensity to agree with the observed histograms.  The necessary shift was consistent in the five fields analyzed.  They interpreted this shift as the CIB, and obtained average values  for the five fields consistent with the values found by Gorjian et al.\ (2000) (Table~1).  They noted that the histogram method is statistically more powerful for finding a real residual and less subject to systematic errors in the star count model than the subtraction approach used by 
Arendt et al.\ (1998).  

Wright (2000) considerably strengthened the case for detection of the CIB at 2.2~$\mu$m.  He used the newly released 2MASS catalog to remove the contribution of Galactic stars brighter than 14th magnitude from the DIRBE maps at 1.25 and 2.2~$\mu$m in 4 dark regions in the North and South galactic polar caps.  Each region was about 2$^\circ$ in radius.  Using the same IPD model as Gorjian et al.\ (2000), Wright obtained 2.2~$\mu$m residuals in his 4 fields consistent with  each other, and consistent with those of Gorjian et al.\ and Wright \& Reese (2000).  Hence, there is a strong case for detection of an isotropic CIB at 2.2~$\mu$m.  The average value of the 2.2~$\mu$m CIB determinations by these three methods is $\nu I_\nu = 21.8 \pm  5.5 $ \mbox{nW m$^{-2}$ sr$^{-1}$}.  The scatter in the 1.25~$\mu$m residuals was too large to claim a detection.  Wright's analysis does provide the strongest current upper limit on the CIB at that wavelength.  The dominant uncertainty in all of these analyses of the near infrared CIB remains the uncertainty in the zodiacal light contribution.

Kashlinsky \& Odenwald (2000) found that the amplitude of the fluctuations in the DIRBE 1.25--4.9~$\mu$m maps varied with $\csc(b)$.  Extrapolating to $\csc(b)$=0, they found positive intercepts which they attributed to fluctuations in the CIB.  They reported rms fluctuations of $15.5 {+3.7\atop -7.0}$, 
$5.9 {+1.6\atop -3.7}$, $2.4 {+0.5\atop -0.9}$, and $2.0 {+0.25 \atop -0.5}$ at 1.25, 2.2, 3.5, and 4.9~$\mu$m respectively, where the errors are 92\% confidence limits.  Adopting their argument that the fluctuations
are expected to be 5--10\% of the CIB, these results are generally consistent with the direct background determinations listed in Table~1.

\section{Summary and Implications}
To put the results of the COBE CIB measurements in a broader perspective, Figure~2 shows the extragalactic background light from the UV to submillimeter wavelengths.  All of the measurements are shown with 2$\sigma$ error bars.  Measurements based upon COBE data are as shown in Figure~1 (see Table~1 for references).  The recently reported CIB measurement from the IRTS mission (Matsumoto et al.\ 2000) is shown as a curve from 1.4--4.0~$\mu$m.  The detections at UV--optical wavelengths are from the recent report by Bernstein (1999) 
\mbox{($\nu I_\nu$ = 12, 15, and 18 nW m$^{-2}$ sr$^{-1}$} at 0.3, 0.55 and 0.8~$\mu$m respectively).  The lower limits to the background light from UV to 2.2~$\mu$m are obtained from integrated galaxy counts obtained from the ground and the Hubble Space Telescope (summarized by Madau \& Pozzetti 2000).  Lower limits in the mid and far infrared are based upon galaxy counts 
with the ISOCAM (7--15~$\mu$m) and ISOPHOT (90--175~$\mu$m) instruments on the ISO mission (Altieri et al.\ 1999, 7 and 
15~$\mu$m; Clements et al.\ 1999, 12~$\mu$m; Juvela, Mattila, \& Lemke 2000, 90 and 150~$\mu$m; Puget et al.\ 1999, 
175~$\mu$m).
The lower limit at 850~$\mu$m is based on the integrated light from source counts with the SCUBA instrument 
(Blain et al.\ 1999).  The shaded region in Figure~2 indicates 2$\sigma$ upper and lower limits on the brightness of the extragalactic background light based upon all of these data.

\begin{figure}
\plotone{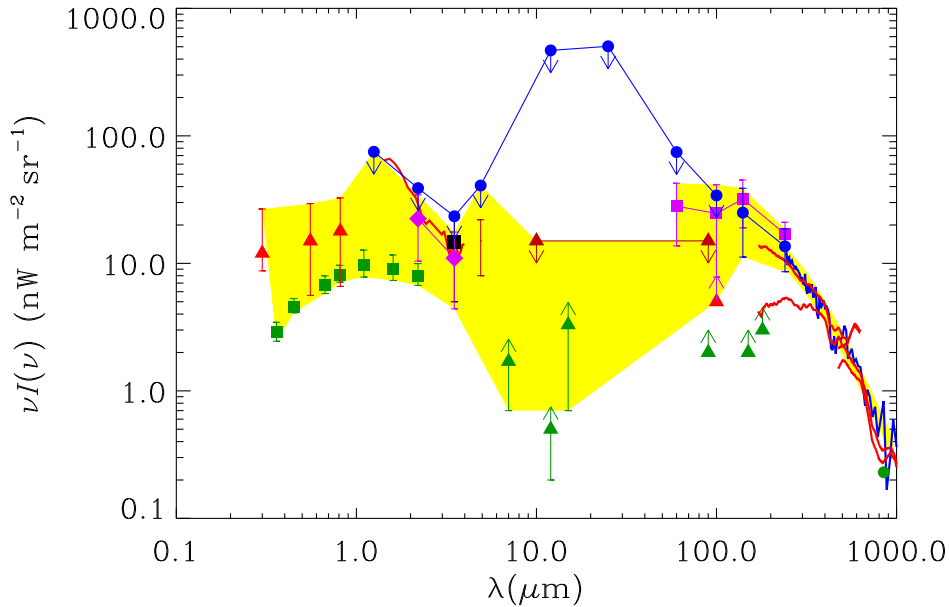}
\caption{Measurements of the extragalactic background light.  See text for references.}
\end{figure}

Figure~2 shows that the resolved component of the extragalactic light is beginning to approach the detected background at the shortest and longest wavelengths in this range.  This reflects a dramatic change in observational knowledge over the past few years.  It is reassuring that the resolved component has not exceeded the claimed CIB detections or upper limits at any wavelength.  The CIB at $\lambda\,\geq\,140\ \mu$m and in the near infrared at 2.2 and 3.5~$\mu$m is  relatively well-determined by the direct sky brightness analyses.  However, systematic uncertainties in the foregrounds are substantial, with none of the CIB detections being positive by more than several times the systematic uncertainties.   Isotropy tests, if done at all, are generally over very limited sky areas.  It would therefore seem premature to consider the differences between detections and lower limits as clear evidence for missing components of the background.  In terms of energy distribution, the $\nu I_\nu$ values in the shortest and longest wavelength regions are rather comparable.  

In the range from 4.9 to 100~$\mu$m the picture is less certain.  Limits from the direct brightness analyses are relatively high, and quite possibly strongly contaminated by residual emission from the interplanetary dust.  The upper limits from fluctuation analyses suggested by Kashlinsky et al.\ (1996b) and Kashlinsky \& Odenwald (2000) in this range are certainly the most restrictive, though dependent upon their estimate of the relationship between fluctuations and total light.  Other fairly restrictive upper limits in this spectral range come from analysis of attenuation of TeV $\gamma$-rays, as reviewed elsewhere at this conference (Stecker 2001).  

Taking the most restrictive limits at face value, the emerging picture is one of substantial background energy in the far infrared and submillimeter, with a comparable level in the near infrared.  To make this picture somewhat quantitative, one can integrate the background intensity in the near infrared (1.25--3.5~$\mu$m), thermal infrared (3.5--100~$\mu$m), and submillimeter (100--2000~$\mu$m) ranges.    One finds integrated intensities of $\sim$ 20, $<56$, and $\sim$ 16 \mbox{nW m$^{-2}$ sr$^{-1}$} respectively.  Using the measurements reported by Bernstein (1999), the UV-optical background 
(0.3--1.25~$\mu$m) contains $\sim$ 25 
\mbox{nW m$^{-2}$ sr$^{-1}$} of additional background energy.  Thus, 0.4 $\leq$ I$_{FIR}$/I$_{UV-OPT}\, \leq$ 1.6, where  I$_{FIR}$ is the integrated long-wavelength background (3.5--2000~$\mu$m) and I$_{UV-OPT}$ is the integrated short-wavelength background (0.3--3.5~$\mu$m).  The long-wavelength and short-wavelength backgrounds contain comparable total energy.

Assuming that the short-wavelength energy comes primarily from redshifted starlight and that the long wavelength radiation comes from starlight absorbed and reradiated by dust, the large far-infrared background implies that a substantial amount of cosmic star formation has been enshrouded in dust, a conclusion reached by many investigators as soon as evidence for the CIB emerged (e.g., Puget et al.\ 1996; Schlegel et al.\ 1998; Hauser et al.\ 1998).  That interpretation has been reinforced by the deep galaxy counts at far infrared wavelengths with the ISO and SCUBA instruments.  The total background energy in the spectral range 0.3--2000~$\mu$m, I$_{EBL}$, is given by I$_{EBL}\,\approx\,$60--120 
\mbox{nW m$^{-2}$ sr$^{-1}$}, or, in terms of the critical density, 
\begin{displaymath}
\Omega_{EBL}\ \approx\ (1.5-3)\times 10^{-6}\, h^{-2},
\end{displaymath}
where \mbox{$h\, =\, \mathrm{H_0}$/(100 km sec$^{-1}$ Mpc$^{-1}$}) and $\mathrm{H_0}$ is the Hubble constant.   For comparison, the integrated EBL contains only about 10\% of the integrated energy in the cosmic microwave background, 
$\Omega_{CMB}\ = 2.5\times 10^{-5}\,h^{-2}$.

With the assumption that the integrated extragalactic background energy primarily arises from nucleosynthesis in stars, one can estimate the mass density consumed in the production of helium and heavier elements, $\rho_Z$, from the relation (Peebles 1993):

\begin{displaymath}
I_{EBL}  =  \left({c\over 4\pi}\right){ 0.007 \rho_Z\,c^2\over 1+z_e}, 
\end{displaymath}
Expressing $\rho_Z$ as a fraction of the critical mass density, $\rho_c$, this yields
\begin{displaymath} 
\Omega_{Z}\ =\ {\rho_Z\over \rho_c}\ = \Omega_{EBL}\left({1+z_e}\over 0.007\right)\ \approx\ (4.3-8.6)\times 10^{-4}\, 
h^{-2},
\end{displaymath}
where we have assumed that most of the energy release occurs at redshift $z_e$ $\approx\ $1.  Since Big Bang nucleosynthesis arguments give a cosmic baryon mass density (Kolb \& Turner 1990; Steigman et al.\ 1999)
\begin{displaymath}
\Omega_{BBN}\ \approx\ (1.1-3.8)\times 10^{-2}\,h^{-2}, 
\end{displaymath}
this simple argument suggests that the current estimate of the integrated extragalactic background light implies conversion of 1--8\% of the cosmic hydrogen into helium and heavier elements by stars.

The present knowledge of the CIB provides significant constraints on models of star formation and galactic evolution (e.g., Dwek et al.\ 1998; Pei, Fall, \& Hauser 1999).   These topics are addressed by others at this conference.

\section{Conclusions}
The extensive and high quality absolute photometric data from the COBE DIRBE and FIRAS instruments have permitted the first definitive measurements of the cosmic infrared background radiation in both the near and far infrared and submillimeter spectral ranges.  The robustness of these determinations is still modest, limited by the systematic uncertainties associated with discriminating strong foreground radiations from the solar system and Galaxy.  In particular, the uncertainty in the amount of scattered and emitted radiation from interplanetary dust is a major remaining source of uncertainty from 
1--100~$\mu$m wavelength.  Significant reduction in that uncertainty may require measurements with instruments in deep space, either out of the ecliptic plane or at several AU from the Sun.  The spectral energy distribution of the CIB sugggests a dominant contribution from starlight, with redshifted UV-optical rest-frame emission of about 45 \mbox{nW m$^{-2}$ sr$^{-1}$} observed in the 0.3--3.5~$\mu$m range, and dust-absorbed and re-emitted energy of about 16 \mbox{nW m$^{-2}$ sr$^{-1}$} observed at 125--2000~$\mu$m.  Present limits allow the possibility of significant additional energy in the 3.5--125~$\mu$m range ($<$ 60 \mbox{nW m$^{-2}$ sr$^{-1}$}).  Assuming that star formation is the major source of the observed background, the large amount of energy in the long wavelength background suggests that a substantial amount of star formation activity at early times was embedded in dust, and that a significant fraction of cosmic baryons has been processed in stars.   The present observations indicate that the energy density in the 0.3--2000~$\mu$m background is about 10\% of that in the cosmic microwave background.

\acknowledgments
The author gratefully acknowledges the assistance of Eli Dwek in the preparation of this paper.  This work was partially supported by NASA grant NAG5--3899 and by NASA contract NAS 5-26555 to the Association of Universities for Research in Astronomy, Inc.

\vspace{5mm}
\centerline{Discussion}
\vspace{5mm}

Michael Werner:  Could you comment on the suggestion of an increased contribution to the galactic FIR background from the ionized medium proposed earlier by Boulanger?

Michael Hauser:  1.  The analysis of Fixsen et al. (1998) included a method which fitted the data to a component of neutral gas (represented by HI and a quadratic term in HI) and a component tracing ionized gas (represented by the FIRAS CII 158\,$\mu$m map).  Their extracted IR background from this method was consistent with that from several other methods.  2.  The DIRBE team analysis of the 140 and 240\,$\mu$m backgrounds was done on dark fields where strong limits could be placed on IR emission from ionized gas. 

Floyd Stecker:  I would point out that the best upper limit we have in the mid-infrared at $\sim 20\,\mu$m is \mbox{$\sim4$\,nW\,m$^{-2}$\,sr$^{-1}$}, which is obtained from the TeV $\gamma$-ray data (Stecker \& De Jager, 1997 -- see Stecker's article, this volume, for the reference).

Hauser:  I am aware of $\gamma$-ray limits, but chose not to address them in my limited time since the subject will be addressed by subsequent speakers.
\end{document}